\documentclass[aps,prd,amsfonts,amsmath,amssymb,preprintnumbers,%
groupedaddress,showpacs,showkeys]{revtex4}

\usepackage{epsfig}
\usepackage{graphicx}
\usepackage{amsmath}
\usepackage{amssymb}
\usepackage{amsfonts,tabularx}

\newcommand{\seq}{\begin{subequations}}
\newcommand{\sen}{\end{subequations}}
\newcommand{\beaa}{\begin{eqnarray*}} 
\newcommand{\enaa}{\end{eqnarray*}}
\newcommand{\bea}{\begin{eqnarray}}
\newcommand{\ena}{\end{eqnarray}}
\newcommand{\be}{\begin{eqnarray}} 
\newcommand{\eq}{\begin{eqnarray}} 
\newcommand{\en}{\end{eqnarray}}

\newcommand{\gH}{\stackrel{\!\!\!\!\circ}{g_H}} 
\newcommand{\gf}{\stackrel{\!\!\!\!\!\circ}{g_{f_0}}} 
\newcommand{\gDs}{\stackrel{\!\!\!\!\!\!\!\!\!\circ}{g_{D_{s0}^\ast}}}
\newcommand{\gBs}{\stackrel{\!\!\!\!\!\!\!\!\circ}{g_{B_{s0}^\ast}}}
\newcommand{\gDa}{\stackrel{\!\!\!\!\!\!\!\!\!\circ}{g_{D_{s1}}}}
\newcommand{\gBa}{\stackrel{\!\!\!\!\!\!\!\!\!\circ}{g_{B_{s1}}}}

\begin{document}

\title{Weak decays of heavy hadron molecules involving 
the ${\mathbf{f_0(980)}}$} 

\author{
Tanja Branz, 
Thomas Gutsche, 
Valery E. Lyubovitskij
\footnote{On leave of absence from the
Department of Physics, Tomsk State University,
634050 Tomsk, Russia} 
\vspace*{1.2\baselineskip}}

\affiliation{Institut f\"ur Theoretische Physik,
Universit\"at T\"ubingen,
\\ Auf der Morgenstelle 14, D-72076 T\"ubingen, Germany
\vspace*{0.3\baselineskip}\\}

\date{\today}

\begin{abstract} 

We study weak decays of the charm- and bottom-strange 
mesons $D_{s0}^\ast(2317)$, $D_{s1}(2460)$, $B_{s0}^\ast(5725)$ 
and $B_{s1}(5778)$ with $f_0(980)$ in the final state by assuming 
a hadronic molecule interpretation for their structures. 
Since in the proposed framework the initial and final states are 
occupied by hadronic molecules, the predictions for observables can 
provide a sensitive tool to further test the hadronic molecule 
structure in future experiments.

\end{abstract} 



\maketitle

\newpage

\section{Introduction}

Over the last decades it became clear that the meson mass spectrum shows 
a much richer structure than one might expect from the conventional 
constituent quark model assigning mesons as $q\bar q$ states. 
For example, the structure of the light scalar mesons below 1 GeV such 
as the $f_0(980)$ have been in the focus. The strong and electromagnetic 
decay properties of the scalar $f_0$ have been intensely studied in 
various models ranging from quarkonium and hybrid structures to compact 
tetraquarks and hadronic molecules (for overview see e.g.  
Ref.~\cite{Klempt:2007cp}). 

Newer experiments delivering data in the heavier mass region also attracted 
interest on mesons with open and hidden charm flavor configurations. Within 
this context one has to mention the $D_{s0}^\ast(2317)$ which has the 
favored spin-parity assignments $J^P=0^+$ and which was first observed by 
{\it BABAR} at SLAC~\cite{Aubert:2003fg}. Shortly afterwards the CLEO 
collaboration~\cite{Besson:2003cp} published their data on the axial 
$D_{s1}(2460)$. Both resonances have been confirmed by 
Belle~\cite{Abe:2003jk}. Up to now the structure issue of the 
$D_{s0}^\ast(2317)$ and $D_{s1}(2460)$ remains an open question. 
Both mesons have therefore been discussed within various structure 
assumptions and theoretical 
frameworks~\cite{Barnes:2003dj,vanBeveren:2003kd,Cheng:2003kg,%
Godfrey:2003kg,Colangelo:2003vg,Bardeen:2003kt,Kolomeitsev:2003ac,%
Fayyazuddin:2003dp,Ishida:2003gu,Azimov:2004xk,%
Colangelo:2004vu,Mehen:2004uj,Hayashigaki:2004st,%
Colangelo:2005hv,Close:2005se,Wei:2005ag,Lu:2006ry,%
Rosner:2006vc,Swanson:2006st,Guo:2006fu,Liu:2006jx,Wang:2006mf,%
Gamermann:2007bm,Lutz:2007sk,Guo:2007up,Guo:2008gp,%
Faessler:2007us,Faessler:2007gv,Faessler:2007hm,Faessler:2008vc}. 

Since their masses are located slightly below the $DK$ and $D^\ast K$ 
thresholds, the $D_{s0}^\ast$ and $D_{s1}$ mesons are clear candidates 
for hadronic molecules with the configurations $D_{s0}^\ast(2317) = DK$ 
and $D_{s1}(2460) = D^\ast K$. In addition, extending this interpretation 
to the bottom sector, the scalar and axial-vector mesons $B^\ast_{s0}(5725)$ 
and $B_{s1}(5778)$ are treated as the equivalents to the charm-strange 
mesons $D_{s0}^\ast(2317)$ and $D_{s1}(2460)$. The bottom-strange  
counterparts $B_{s0}^\ast(5725)$ and $B_{s1}(5778)$ are consequently 
also described as bound states with $B_{s0}^\ast(5725)=B\bar K$ and 
$B_{s1}(5778)=D^\ast K$. The decay properties of these hadronic molecules 
were studied within the same effective Lagrangian 
approach~\cite{Faessler:2007us,Faessler:2007gv,Faessler:2007hm,%
Faessler:2008vc,Dong:2008gb,Branz:2007xp,Branz:2008ha}. 
Within this covariant model for hadronic bound states, the molecular 
structure is considered by the compositeness condition 
$Z = 0$~\cite{Weinberg:1962hj,Salam:1962ap,Efimov:1993zg,%
Faessler:2007us,Faessler:2007gv,Faessler:2007hm,%
Faessler:2008vc,Dong:2008gb,Branz:2007xp,Branz:2008ha,%
Ivanov:1996pz,Baru:2003qq,Hanhart:2007wa} 
which implies that the renormalization constant of the hadronic molecule 
field is set equal to zero. The composite object therefore exists 
exclusively as a bound state of its constituents. This condition also 
provides a method to fix the coupling between the hadronic molecule and 
its constituent mesons in a self-consistent way. Furthermore, our 
theoretical framework also features finite size effects of the meson 
molecules controlled by size parameters which are the only adaptive variables.

In the present paper the $f_0(980)$ properties are studied in weak hadronic 
decays of the scalar $D_{s0}^\ast(2317)$ and its bottom-strange counterpart 
$B_{s0}^\ast(5725)$ as well as in the weak non-leptonic decay processes of 
the axial-vector mesons $D_{s1}(2460)$ and $B_{s1}(5778)$. Since we deal 
with transition processes between hadronic molecules, the decay properties 
involve twice the effect of meson bound states: In the initial heavy meson 
system and in the final scalar $f_0$. For this reason the results might 
provide a sensitive observable to test the issue of hadronic molecule 
structure accessible in future experiments.

The paper is organized as follows. In the next section \ref{sec:1} we give 
a short introduction to the effective Lagrangian approach we use for the 
description of hadronic bound states. In section \ref{sec:2} we deal with 
the weak non-leptonic decays of the scalar mesons $D_{s0}^\ast(2317)$ and 
$B_{s0}^\ast(5725)$, where the meson molecule $f_0$ appears in the final 
state. The $D^\ast K\pi$ coupling $g_\pi$, which we need for the 
$D_{s1}^+(2460)\to f_0\pi^+$ transition, is derived in Sec.~\ref{sec:2} 
from the $D_{s}\to \pi f_0$ decay. Thereby we also obtain the 
$D^\ast\to K\pi$ decay width as a byproduct of our analysis. 
In Sec.~\ref{sec:3} we finally compute the $f_0$-production in hadronic 
decays of the axial-vector mesons $D_{s1}(2460)$ and $B_{s1}(5778)$.

\section{Basics of the model}\label{sec:1}

An assortment of mesons with masses lying close to two-body thresholds 
are good candidates for mesonic bound states and have therefore been 
studied assuming a hadronic molecule structure. For instance, in 
Refs.~\cite{Faessler:2007us,Faessler:2007gv,Faessler:2007hm,%
Faessler:2008vc,Dong:2008gb,Branz:2007xp,Branz:2008ha} we developed 
a field-theoretical approach to study the properties of hadronic molecules 
($f_0(980)$, $D_{s0}^\ast(2317)$, $D_{s1}(2460)$, $B_{s0}^\ast(5725)$,  
$B_{s1}(5778)$ and $X(3872)$) as bound states of two mesons. 
Since above states are close to the corresponding thresholds, we used 
the following dominant composite structures: 
\eq 
\big|f_0\big> &=& 
\frac{1}{\sqrt2}\big(\big|K^+K^-\big>+\big|K^0\bar K^0\big>\big)\,, 
\nonumber\\
\big|D_{s0}^{\ast\,+}\big> &=& 
\frac{1}{\sqrt2}\big(\big|D^+K^0\big>+\big|D^0 K^+\big>\big)\,, 
\nonumber\\ 
\big|D_{s1}^+\big>&=&\frac{1}{\sqrt2}\big(\big|D^{\ast\,+}K^0\big>
+\big|D^{\ast\,0}K^+\big>\big)\,, \\  
\big|B_{s0}^{\ast\,0}\big> &=& 
\frac{1}{\sqrt2}\big(\big|B^+K^-\big>+\big|B^0 \bar K^0\big>\big)\,, 
\nonumber\\ 
\big|B_{s1}^0\big>&=&\frac{1}{\sqrt2}\big(\big|B^{\ast\,+}K^-\big>
+\big|B^{\ast\,0}\bar K^0\big>\big)\,. \nonumber 
\en 
The model for hadronic molecules 
$H = f_0(980)$, $D_{s0}^\ast(2317)$, $D_{s1}(2460)$, $B_{s0}^\ast(5725)$ 
composed of two meson constituents 
$M_1$ and $M_2$ is thereby based on the nonlocal interaction Lagrangians 
\eq\label{Lagrangian_HM1M2}
{\cal L}_{HM_1M_2}&=& g_{H} H(x) \int dy 
\,\Phi_H(y^2) \, M^T_1(x+w_{21}y) \, M_2(x-w_{12}y) + \text{H.c.}\,, 
\en 
where $M_1$ and $M_2$ are the doublets of the meson fields: 
\eq 
K=\left(\begin{array}{c}K^+\\K^0\end{array}\right)\,, \ \  
D=\left(\begin{array}{c}D^{0}\\D^{+}\end{array}\right)\,, \ \ 
D^\ast_\mu=\left(\begin{array}{c}D^{\ast\,0}\\D^{\ast\,+}
\end{array}\right)_\mu\,, 
\ \ 
B=\left(\begin{array}{c}B^+\\B^0\end{array}\right)\,, \ \ 
B^\ast_\mu=\left(\begin{array}{c}B^{\ast\,+}\\B^{\ast\,0}
\end{array}\right)_\mu\, 
\en 
and their antiparticles. The symbol $T$ refers to the transpose of $M_1$. 
The kinematic variable $w_{ij}$ is defined by $w_{ij} = m_i/(m_i + m_j)$ 
where $m_1$ and $m_2$ are the masses of $M_1$ and $M_2$.

The finite size of the hadronic molecule is introduced through the correlation 
function $\Phi_H(y^2)$ which describes the distribution of the 
constituent mesons. Its Fourier transform $\widetilde\Phi_H(k_E^2)$ appears as 
the form factor in our calculations, where, in the present analysis, 
we have chosen a Gaussian form 
\eq 
\widetilde\Phi_H(k_E^2)=\exp(-k_E^2/\Lambda_H^2) 
\en 
in Euclidean momentum space. The size parameter $\Lambda_H$ controls 
the spatial extension of the hadronic molecule and is varied between 
1 - 2 GeV. The local case (LC), describing point-like interaction, is 
defined for $\Lambda_H\to \infty$. (Note this limit can be applied to 
convergent matrix elements only). The size parameters $\Lambda_H$ 
are the only adjustable parameters in our framework.

\begin{figure}[htbp]  
\includegraphics[trim= 0.5cm 22.5cm 0.0cm 1.50cm, clip,scale=0.6]{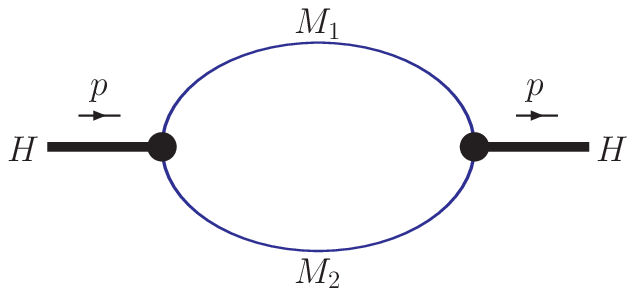} 
\caption{Mass operator of the hadronic molecule.}
\label{fig1}
\end{figure}

The coupling constants between the hadronic molecules and its building 
blocks, the constituent mesons, are fixed self-consistently by the 
compositeness 
condition~\cite{Weinberg:1962hj,Salam:1962ap,Efimov:1993zg,%
Faessler:2007us,Faessler:2007gv,Faessler:2007hm,Faessler:2008vc,%
Dong:2008gb,Branz:2007xp,Branz:2008ha,Ivanov:1996pz,Baru:2003qq,%
Hanhart:2007wa}. 
The dynamics of the bound state is therefore related to its constituents 
by setting the field renormalization constant to zero. Because of this 
constraint, the coupling constants are no input parameters but are fixed 
within this theoretical framework. The number of free 
variables is therefore reduced to the size parameters $\Lambda_H$. 
For the generic hadronic molecule $H=(M_1M_2)$, the compositeness condition 
is given by the relation 
\eq 
Z_{H} = 1 - \Sigma^\prime_H(m_{H}^2) = 0\,,
\en 
where $\Sigma^\prime_H(m_{H}^2) = g_{H}^2\Pi^\prime_H(m_{H}^2)$ 
is the derivative of the mass operator (see Fig.~\ref{fig1}) and 
$m_H$ is the mass of hadronic molecule. 

In the mesonic molecule picture all decays proceed via intermediate 
states which are the composite mesons of the hadronic bound state. 
We describe the dynamics of the intermediate states by free propagators 
given by the standard expressions
\eq 
iS_M(x-y)=\left<0|TM(x)M^\dagger (y)|0\right>=\int\frac{d^4k}{(2\pi)^4i}\, 
e^{-ik(x-y)} S_M(k),\quad S_M(k)=
\frac{1}{m_M^2-k^2-i\epsilon}
\en 
for pseudoscalar and scalar fields $M$ and by
\eq
iS_{M^\ast}^{\mu\nu}(x-y)=
\left<0|TM^{\ast\,\mu}(x)M^{\ast\,\nu\,\dagger} (y)|0\right>=
\int\frac{d^4k}{(2\pi)^4i}\,e^{-ik(x-y)} S^{\mu\nu}_{M^\ast}(k)\,,
\quad S^{\mu\nu}_{M^\ast}(k)=
\frac{-g^{\mu\nu}+k^\mu k^\nu/m_{M^\ast}^2}{m_{M^\ast}^2-k^2-i\epsilon}
\en 
in case of vector and axial-vector fields $M^\ast$.

\def\arraystretch{1.5} 
For the $D$ and $B$ meson masses we use the the values 
quoted in~\cite{Amsler:2008zz} and estimated in~\cite{Guo:2006fu}:    
\eq 
& &m_{D^+}=1.8696\text{ GeV},    \hspace*{.2cm} 
   m_{D^0}=1.8648\text{ GeV},   \hspace*{.2cm}
   m_{D_s^+}=1.96849\text{ GeV}, \hspace*{.2cm}
   m_{D^{\ast\,+}}=2.01027\text{ GeV}, \hspace*{.2cm}
   m_{D^{\ast\,0}}=2.00697\text{ GeV}, \nonumber\\ 
& &m_{B^+}=5.2791\text{ GeV}, \hspace*{.2cm}
   m_{B^0}=5.2795\text{ GeV}, \hspace*{.2cm}
   m_{B^{\ast\,+}}=5.3251\text{ GeV}, \hspace*{.2cm}
   m_{B^{\ast\,0}}=5.3251\text{ GeV}, \\
& &m_{D_{s0}^\ast}=2.3178\text{ GeV}, \hspace*{.2cm}
   m_{B_{s0}^\ast}=5.725\text{ GeV},  \hspace*{.2cm}
   m_{D_{s1}}=2.4596\text{ GeV},      \hspace*{.4cm}  
   m_{B_{s_1}}=5.778\text{ GeV}. \nonumber 
\en 
Below we list our previous predictions for the couplings $g_H$ obtained 
for the respective molecular states. In particular, for the 
$f K\bar K$-coupling we obtained~\cite{Branz:2007xp} 
\eq\label{couplings1}  
g_{f_0}=3.09 \text{ GeV}\quad(\Lambda_{f_0}=1\text{ GeV})\,,\quad 
g_{f_0}=2.9  \text{ GeV}\quad(\text{LC})\,. 
\en 
The coupling constants of the $D_{s0}^\ast$ and $D_{s1}$ 
mesons have already been calculated 
in~\cite{Faessler:2007us,Faessler:2007gv,Faessler:2007hm}:   
\eq\label{couplings2} 
g_{D_{s0}^\ast}&=11.26\text{ GeV}
\quad(\Lambda_{D_{s0}^\ast}=1\text{ GeV}),\, 
\hspace*{.7cm} 
g_{D_{s0}^\ast}&=9.9\text{ GeV}
\quad(\Lambda_{D_{s0}^\ast}=2\text{ GeV}),
\hspace*{.7cm} 
g_{D_{s0}^\ast}=8.98\text{ GeV}\quad(\text{LC})\,,  
\nonumber\\ 
g_{D_{s1}}&=11.62\text{ GeV}
\quad(\Lambda_{D_{s1}}=1\text{ GeV})\,,  
\hspace*{.7cm}
g_{D_{s1}}&=10.17\text{ GeV}\quad(\Lambda_{D_{s1}}=2\text{ GeV})\,.
\en 
The results for the couplings of the $B_{s0}^\ast$ and $B_{s1}$ mesons 
to their constituents for different size parameters $\Lambda$ 
are~\cite{Faessler:2008vc}: 
\eq 
& &g_{B^\ast_{s0}}=27.17\text{ GeV}
\quad(\Lambda_{B^\ast_{s0}}=1\text{ GeV}),
\quad  
g_{B^\ast_{s0}}=23.21\text{ GeV}
\quad(\Lambda_{B^\ast_{s0}}=2\text{ GeV}), 
\quad 
g_{B^\ast_{s0}}=20.10\text{ GeV}\quad\text{(LC)}\,,  
\nonumber\\ 
& &g_{B_{s1}}=25.64\text{ GeV}\quad(\Lambda_{B_{s1}}=1\text{ GeV}), 
\quad 
g_{B_{s1}}=22.14\text{ GeV}\quad(\Lambda_{B_{s1}}=2\text{ GeV})\,. 
\en 
One should stress that the coupling constants $g_{f_0}$, $g_{_{D_{s0}^\ast}}$ 
and $g_{_{B_{s0}^\ast}}$ of the scalar mesons $f_0$, $D_{s0}^\ast$, 
and $B_{s0}^\ast$ remain finite when we remove the cutoff 
$\Lambda_H \to \infty$. For the axial mesons $D_{s1}$ and $B_{s1}$ 
the couplings $g_{_{D_{s1}}}$ and $g_{_{B_{s1}}}$ are finite in the 
local limit when we neglect the longitudinal part $k^\mu k^\nu/m_{M^\ast}^2$ 
of the constituent vector meson propagator. In this case all the couplings 
are given analytically by 
\eq
\frac{1}{g_H^2} = \frac{2}{(4 \pi m_H)^2} \,
\biggl\{ \frac{m_1^2 - m_2^2}{m_H^2} \,
{\rm ln}\frac{m_1}{m_2} \, - 1 \, + \,
\frac{m_H^2 (m_1^2+m_2^2)
- (m_1^2 - m_2^2)^2}{m_H^2\sqrt{-\lambda}} \,
\sum\limits_\pm {\rm arctan}\frac{z_\pm}{\sqrt{-\lambda}} \biggr\}
\en
where $z_\pm = m_H^2 \pm (m_1^2 - m_2^2)$ and
\eq\label{Kaellen}
\lambda \doteq \lambda(m_H^2,m_1^2,m_2^2) =
m_H^4 + m_1^4 + m_2^4
- 2 m_H^2 m_1^2
- 2 m_H^2 m_2^2
- 2 m_1^2 m_2^2
\en
is the K\"allen function. When writing the mass $m_H$ of the hadronic 
molecule in the form $m_H = m_1 + m_2 - \epsilon\,,$ 
where $\epsilon$ represents the binding energy, we can perform 
an expansion of $g_H^2$ in powers of $\epsilon$. The leading-order 
${\cal O}(\sqrt{\epsilon})$ result 
\eq\label{eq:g} 
\frac{\gH^{\!\!\! 2}}{4\pi} 
= \frac{(m_1+m_2)^{5/2}}{\sqrt{m_1m_2}} \, \sqrt{8\epsilon} 
\en 
in agreement with the one derived in 
Refs.~\cite{Guo:2008zg,Weinberg:1962hj,Baru:2003qq,Hanhart:2007wa} 
based on a formalism which also used the compositeness condition $Z_H=0$. 

Numerical results for the coupling constants $\gH$ 
\eq 
\gf = 2.74\text{ GeV}\,,  \hspace*{.4cm} 
\gDs = 8.27\text{ GeV}\,, \hspace*{.4cm} 
\gDa = 8.63\text{ GeV}\,, \hspace*{.4cm} 
\gBs = 19.63\text{ GeV}\,, \hspace*{.4cm} 
\gBa = 19.01\text{ GeV}\,. 
\en 
compare well with the results obtained in the local case without 
the $\epsilon$ expansion and in the nonlocal case 
(see Eqs.~(\ref{couplings1}) and (\ref{couplings2})). Note that 
in the calculation of $\gf$ we use the averaged kaon mass 
$\bar m_K = (m_{K^\pm} + m_{K^0})/2$.    

For consistency we also analyze the couplings $g_H$ 
and $\gH$ in the heavy 
quark limit (HQL), where the masses of the heavy mesons together with the 
heavy quark masses go to infinity. The scaling of the coupling 
constant $g_{D_{s0}^\ast}$ in the HQL was already discussed in~\cite{Faessler:2007gv}. 
It was shown that $g_{D_{s0}^\ast}$, both for the nonlocal and the local case, is proportional 
to the charm quark mass or the mass of the $D_{s0}^\ast$ meson (see Eqs.(57) and (58) of Ref.~\cite{Faessler:2007gv}). This result 
is simply extended to the cases of the $B_{s0}^\ast$ coupling and 
of the couplings of the axial states $D_{s1}$ and $B_{s1}$. In particular, 
for the nonlocal case the result for $g_H$ in the HQL is:
\eq\label{HQL_nonl}
\frac{1}{g_H^2} &=&
\frac{1}{(4 \pi m_H)^2} \, \int\limits_0^\infty
\frac{\displaystyle{d\alpha}}
{\displaystyle{1 + \mu_K^2 \alpha}} \,\,
\tilde\Phi^2_H(\alpha) \,, 
\en 
where $\mu_K = m_K/\Lambda_H$.  
In the local case the HQL reads as:
\eq\label{HQL_loc}
\frac{1}{g_H^2} =
\frac{1}{(4 \pi m_H)^2} \, {\rm ln}\frac{m_H^2}{m_K^2} \,.
\en 
Hence, the coupling of the heavy-light
molecules to the constituents is proportional to the heavy quark mass 
(or the molecule mass $m_H = m_Q + {\cal O}(1)$). 
Therefore, we deduce the following 
relations between the coupling constants $g_H$ in the HQL: 
\eq\label{relations_HQL}
& &g_{D_{s0}^\ast} = g_{D_{s1}}\,, \hspace*{.25cm} 
g_{B_{s0}^\ast} = g_{B_{s1}}\,, \nonumber\\
& &\frac{g_{B_{s0}^\ast}}{g_{D_{s0}^\ast}} = \frac{g_{B_{s1}}}{g_{D_{s1}}}  
\simeq \frac{m_{B_{s0}^\ast}}{m_{D_{s0}^\ast}} 
\simeq \frac{m_{B_{s1}}}{m_{D_{s1}}} \,. 
\en 
This scaling behavior is also evident from Eq.~(\ref{eq:g}), where the couplings $\gH$ 
behave in the HQL as: 
\eq 
\frac{\gH^{\!\!\! 2}}{4\pi} 
= m_H^2 \, \sqrt{\frac{8\epsilon}{m_K}} \,. 
\en 
Keeping in mind that the binding energy $\epsilon$ is 
approximately the same for all four states 
($D_{s0}^\ast$, $B_{s0}^\ast$, $D_{s1}$, $B_{s1}$), we deduce  
that in the HQL the relations (\ref{relations_HQL}) are also valid for the 
leading-order couplings $\gH$. Using the previous numerical values for the $g_H$ and $\gH$ couplings 
one can see that the HQL relations (\ref{relations_HQL}) are 
fulfilled with a good accuracy. It also explains the phenomenon 
that the bottom meson couplings are 2.2 - 2.8  times
larger than the charm ones.

\section{${\mathbf{D_{s0}^\ast(2317)}}$ and ${\mathbf{B_{s0}^\ast(5725)}}$ 
decays}\label{sec:2}

In this section we deal with the $f_0$-production properties in weak 
hadronic decays of the heavy scalar mesons $D_{s0}^\ast(2317)$ and 
$B_{s0}^\ast(5725)$. Here the final states of the 
$D_{s0}^{\ast\,+}\to f_0X$ decay are occupied by the charged mesons 
$X=\pi^+,K^+,\rho^+$ and the scalar $f_0$. The decay pattern of the 
neutral $B_{s0}^{\ast\,0}$ decay is richer and we deal with final 
$\pi^0,\;K^0,\;\rho^0,\;\omega,\;\eta$ and $\eta^\prime$ mesons besides 
the $f_0$.

\begin{figure}[htbp]
\includegraphics[trim= 0cm 16cm 0cm 7.0cm,clip,scale=0.6]{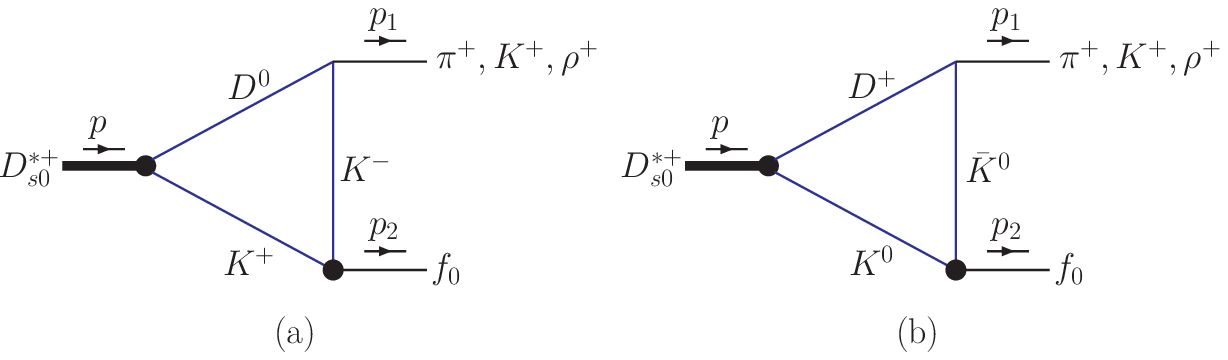}
\caption{Diagrams contributing to the $D_{s0}^{\ast\,+}\to f_0 X$ decays 
with $X=\pi^+,\,K^+$ and $\rho^+$.}
\label{fig2}
\end{figure}

\begin{figure}[htbp]
\includegraphics[trim= 0cm 16cm 0cm 7.0cm,clip,scale=0.6]{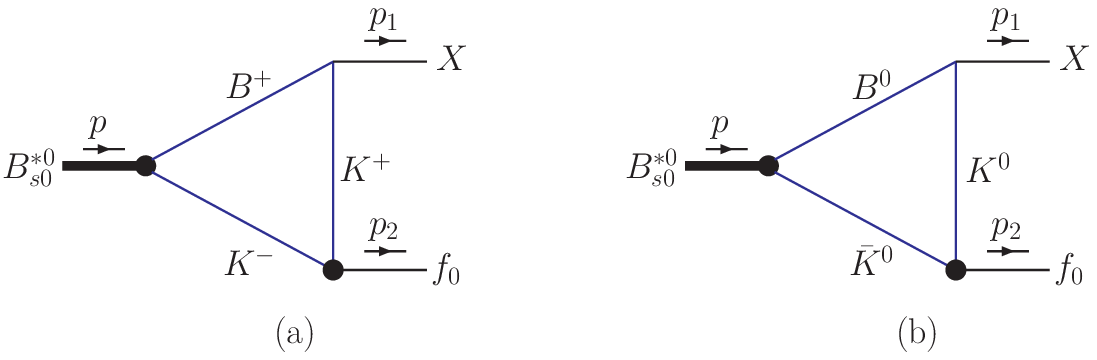}
\caption{Diagrams contributing to the $B_{s0}^{\ast\,0} \to f_0 X$ decays 
with $X=\pi^0,\;K^0,\;\rho^0,\;\omega,\;\eta$ and $\eta^\prime$.}
\label{fig3}
\end{figure}

Since both heavy quark systems are assumed to be of molecular structure 
the decays proceed via intermediate kaons and $D$ or $B$ mesons as 
indicated in the diagrams of Figs. \ref{fig2} and \ref{fig3}.

The couplings of the hadronic molecules to the constituent mesons 
in the loop are fixed by the compositeness condition. The coupling 
constants between the intermediate $K$, $D$ and $B$ mesons and the 
final decay products $\pi,K,\rho,\omega\,\eta$ and $\eta^\prime$ are 
obtained from the $D$ and $B$ meson partial decay widths. The latter 
constants are given by following expressions, where we distinguish 
between final pseudoscalar ($P$) and vector mesons ($V$):
\bea
g_{_{\scriptstyle{HKP}}}^{c(n)}&=&
\sqrt{\frac{16\pi\,\Gamma({H\to K\,P})\,
m_H^3}{\lambda^\frac12(m_H^2,m_K^2,m_P^2)}}\,,
\quad(P=K,\pi,\eta,\eta^\prime,\quad H = D,B)\,,\label{eq:g1}\\
g_{HKV}^{c(n)}&=&\sqrt{\frac{64\pi\,\Gamma({H \to K\,V}) 
\,m_H^3 \,m_V^2}{\lambda^\frac32(m_H^2,m_K^2,m_V^2)}}\,,
\quad(V=\rho,\omega,\quad H = D,B)\,,\label{eq:g2}
\ena
with the K\"allen function $\lambda(x,y,z)$ defined in Eq.~(\ref{Kaellen}).  
The superscript $c$ ($n$) denotes the decays of the charged (neutral) 
$D$ and $B$ mesons.

The couplings governing the $D_{s0}^{\ast}\to f_0 P$ and 
$B_{s0}^{\ast}\to f_0 P$ decays we calculate from
\bea
g_{_{\scriptstyle{D_{s0}^\ast f_0 \,P}}}&=&
\frac{g_{_{D_{s0}^\ast}}g_{f_0}}{(4\pi)^2}
\big[g_{HKP}^cI(m_{D^+}^2,m_{K^0}^2)
+g_{HKP}^nI(m_{D^0}^2,m_{K^+}^2)\big]\,,\\
g_{_{\scriptstyle{B_{s0}^\ast f_0 \,P}}}&=&
\frac{g_{_{B_{s0}^\ast}}g_{f_0}}{(4\pi)^2}
\big[g_{HKP}^c I(m_{D^+}^2,m_{K^+}^2)
+g_{HKP}^nI(m_{D^0}^2,m_{K^0}^2)\big]\,,
\ena
where $I(m_H^2,m_K^2)$ denotes the loop integral 
\eq 
I(m_H^2,m_K^2)=\int\frac{d^4k}{\pi^2i}\,\widetilde \Phi_{f_0}(-k^2)
\widetilde \Phi_{_{\scriptstyle{H_{s0}^\ast}}}
\big(-(k-\frac p2+\omega p_{_{\scriptstyle{H_{s0}^\ast}}})^2\big)
\,S_H\big(k-\frac p2+p_{_{\scriptstyle{H_{s0}^\ast}}}\big)
S_K\big(k-\frac p2\big)S_K\big(k+\frac p2\big)
\,,
\en 
with $H_{s0}^\ast=B_{s0}^{\ast\,0}\,,D_{s0}^{\ast\,+}$. 

The decay widths are finally obtained from
\eq 
\Gamma(H_{s0}^\ast\to f_0 \,P)&=&
\frac{g_{_{\scriptstyle{H_{s0}^\ast f_0\,P}}}^2} 
{16\pi m_{H_{s0}^{\ast}}^3} \, 
\lambda^\frac12(m_{H_{s0}^\ast}^2,m_{f_0}^2,m_{P}^2) 
\,.
\en
For the decays with a final vector meson, 
$D_{s0}^\ast/B_{s0}^\ast\to f_0 V$, we proceed in analogy. 
For simplicity, we restrict in the following to the 
$D_{s0}^{\ast\,+}\to f_0\rho^+$ decay since the corresponding 
expressions for the bottom $B_{s0}^\ast$ decays only differ in 
the masses and couplings, while the structure remains the same.

Again, the Feynman integral
\eq 
I^\mu(m_{D}^2,m_{K}^2)&=&\int\frac{d^4k}{\pi^2i}\,
\widetilde \Phi_{f_0}(-k^2) \, 
\widetilde \Phi_{D_{s0}^\ast}
\big(-(k-\frac p2+\omega p_{D_{s0}^\ast})^2\big) \, 
(2k+p_{D_{s0}^\ast})^\mu \nonumber\\ 
&\times&S_D\big(k-\frac p2+p_{D_{s0}^\ast}\big) 
S_K\big(k-\frac p2\big)S_K\big(k+\frac p2\big)  
\en 
defines the transition matrix element ${\cal M}^\mu$ which is given by
\eq 
{\cal M}^\mu&=&\frac{g_{D_{s0}^\ast}g_{f_0}}{(4\pi)^2}
\big[g_{HK\rho}^cI^\mu(m_{D^+}^2,m_{K^0}^2)
+g_{HK\rho}^nI^\mu(m_{D^0}^2,m_{K^+}^2)\big]\nonumber\\
&=&F_1(m_{D_{s0}^\ast}^2,m_{f_0}^2,m_\rho^2)\,p_f^\mu
+F_2(m_{D_{s0}^\ast}^2,m_{f_0}^2,m_\rho^2)\,p_\rho^\mu\,.
\en 
In the second line ${\cal M}^\mu$ is expressed in terms of the form 
factors $F_1$ and $F_2$ by writing the matrix element as a linear 
combination of the $f_0$ and $\rho$ meson momenta $p_f$ and $p_\rho$. 
We perform this decomposition since the form factor $F_1$ defines the 
coupling constant of the decay
\eq 
F_1(m_{D_{s0}}^2,m_{f_0}^2,m_\rho^2)\equiv g_{D_{s0}^\ast f_0\rho} 
\en 
and therefore characterizes the decay width with
\eq 
\Gamma(D_{s0}^{\ast\, +} \to f_0 \rho^+)=
\frac{g_{D_{s0}^\ast f_0\rho}^2} 
{64\pi m^3_{D_{s0}^\ast}m_\rho^2}\, 
\lambda^\frac32(m_{D_{s0}^\ast}^2,m_{f_0}^2,m_{\rho}^2) 
\,.
\en  

First we indicate the results for the coupling constants at the 
secondary interaction vertex as deduced from the decays $B/D\to KX$ 
($X=\pi,\,K,\eta^\prime,\eta,\omega,\rho$). In Table \ref{tab:3} 
we summarize the branching ratios (Br) as taken from
and the resulting couplings 
$g_X^{c(n)}$ (via Eqs. (\ref{eq:g1}) and (\ref{eq:g2})) involving charged 
($c$) and neutral ($n$) $B$ and $D$ mesons.
\begin{table}[htbp]
\caption{Coupling constants deduced from the decays $B/D\to KX$ 
with $X=\pi,\;K,\;\eta^\prime,\;\eta,\;\omega,\;\rho$.} 
\label{tab:3}
\begin{tabularx}{\textwidth}{|XXX|XXX|}
\hline
Channel&Br~\cite{Eidelman:2004wy,Amsler:2008zz} 
       &$g_{X}^n$&Channel
       &Br~\cite{Eidelman:2004wy,Amsler:2008zz}
       &$g_{X}^c$\\ 
\hline
$D^0\to \pi^+ K^-$&$(3.89\pm0.05)\,\%$&$2.88\cdot 10^{-6}$ GeV
&$D^+\to\pi^+\bar{K^0}$&$(2.83\pm0.18)\,\%$&$0.14\cdot 10^{-5}$ GeV\\
\hline
$D^0\to K^+K^-$&$(3.93\pm0.08)\cdot10^{-3}$&$0.83\cdot 10^{-6}$ GeV
&$D^+\to K^+\bar{K^0}$&$(5.7\pm0.5)\cdot10^{-3}$&$0.63\cdot 10^{-6}$ GeV\\ 
\hline
$D^0\to\rho^+K^-$&$(10.8\pm0.7)\,\%$&$2.92\cdot 10^{-6}$
&$D^+\to\rho^+\bar{K^0}$&$(7.3\pm2.5)\,\%$&$0.15\cdot 10^{-5}$\\ 
\hline\hline
$B^0\to K^0\pi^0$&$(9.8\pm0.6)\cdot 10^{-6}$&$3.36\cdot 10^{-8}$ GeV
&$B^+\to K^+\pi^0$&$(1.29\pm0.06)\cdot 10^{-5}$&$3.73\cdot 10^{-8}$ GeV\\ 
\hline
$B^0\to K^0\eta^\prime$&$(6.5\pm0.4)\cdot 10^{-5}$
&$0.91\cdot 10^{-7}$ GeV&$B^+\to K^+\eta^\prime$
&$(7.02\pm0.25)\cdot 10^{-5}$&$8.84\cdot 10^{-8}$ GeV\\ 
\hline
$B^0\to K^0\eta$&$<1.9\cdot 10^{-6}$&$<0.15\cdot 10^{-7}$ GeV
&$B^+\to K^+\eta$&$(2.7\pm0.9)\cdot 10^{-6}$&$0.17\cdot 10^{-7}$ GeV\\ 
\hline
$B^0\to K^0\bar{K^0}$&$(9.6^{+2.0}_{-1.8})\cdot 10^{-7}$
&$1.06\cdot 10^{-8}$ GeV&$B^+\to K^+\bar{K^0}$
&$(1.36\pm0.27)\cdot 10^{-6}$&$1.22\cdot 10^{-8}$ GeV\\ 
\hline
$B^0\to K^0\omega$&$(5.0\pm0.6)\cdot 10^{-6}$&$1.41\cdot 10^{-9}$
&$B^+\to K^+\omega$&$(6.7\pm0.8)\cdot 10^{-6}$&$1.57\cdot 10^{-9}$\\ 
\hline
$B^0\to K^0\rho^0$&$(5.4\pm0.9)\cdot 10^{-6}$&$0.14\cdot 10^{-8}$
&$B^+\to K^+\rho^0$&$(4.2\pm0.5)\cdot 10^{-6}$&$1.23\cdot 10^{-9}$\\ 
\hline
\end{tabularx}
\end{table}

In Tables \ref{tab:1} and \ref{tab:4} we summarize the results for the 
coupling constants and decay widths of the $D_{s0}^{\ast\,+}(2317)$ and 
$B_{s0}^{\ast\,0}(5725)$ decays. We also indicate the dependence of the 
results for different sets of size parameters $\Lambda_H$.  
Compared to the local case (LC) finite size effects induce a reduction 
of the $D_{s0}^\ast$ decay widths by up to 50$\%$. For the $B^\ast_{s0}$ 
decays inclusion of finite size parameters leads to a reduction of the 
partial decay widths by up to a factor of 10.

For the $D_{s0}^{\ast\,+}$ decays we predict a decay pattern with 
\eq
\Gamma(f_0\rho^+)> 
\Gamma(f_0\pi)>  
\Gamma(f_0K^+)\,, 
\en 
where the decay width of each sequential decay mode is reduced by about 
an order of magnitude. Here we introduce the shortened notation 
$\Gamma(D_{s0}^{\ast\,+}\to H_1 H_2) = \Gamma(H_1 H_2)$. 
In the case of $B_{s0}^{\ast\,0}$ the weak decay mode 
$B_{s0}^{\ast\,0}\to f_0\eta^\prime$ dominates the transitions with 
the decay hierarchy
\eq
\Gamma(f_0\eta^\prime)>\Gamma(f_0\pi)\approx 
\Gamma(f_0\rho)\approx\Gamma(f_0\omega)>
\Gamma(f_0 K)\approx\Gamma(f_0\eta)\,.
\en
\begin{table}[thbp]
\caption{
$D_{s0}^{\ast\,+} \to f_0 X$ decay properties with 
$X = \pi^+,\; K^+,\; \rho^+$.}
\label{tab:1}
\begin{tabularx}{\linewidth}{|l|XX|Xl|XX|}
\hline
&\multicolumn{2}{c|}{$D_{s0}^{\ast\,+}\to f_0\pi^+$}
&\multicolumn{2}{c|}{$D_{s0}^{\ast\,+}\to f_0K^+$}
&\multicolumn{2}{c|}{$D_{s0}^{\ast\,+}\to f_0\rho^+$}\\ 
\hline
$\Lambda_H$ [GeV]&$g_{_{\scriptstyle{D_{s0}^\ast f_0\pi}}}$ [GeV]
&$\Gamma$ [GeV]&$g_{_{\scriptstyle{{D_{s0}^\ast f_0K}}}}$ [GeV]
&$\Gamma$ [GeV]&$g_{_{\scriptstyle{{D_{s0}^\ast f_0\rho}}}}$
&$\Gamma$ [GeV]\\ 
\hline
LC&$1.83\cdot 10^{-6}$  &$2.35\cdot 10^{-14}$  &$6.51\cdot10^{-7}$  
&$2.75\cdot10^{-15}$  &$2.37\cdot 10^{-6}$&$1.60\cdot 10^{-13}$ \\ 
\hline
$\Lambda_{D_{s0}^{\ast}}=2$,\, $\Lambda_{f_0}=1$&$1.34\cdot10^{-6}$   
&$1.26\cdot10^{-14}$  &$4.86\cdot10^{-7}$  &$1.53\cdot10^{-15}$  
&$1.95\cdot 10^{-6}$ &$1.08\cdot 10^{-13}$\\ 
\hline
$\Lambda_{D_{s0}^{\ast}}=1$,\,$\Lambda_{f_0}=1$&$1.28\cdot 10^{-6}$  
&$1.14\cdot 10^{-14}$  &$4.68\cdot10^{-7}$  &$1.42\cdot10^{-15}$  
&$1.98\cdot10^{-6}$&$1.11\cdot10^{-13}$  \\ 
\hline
\end{tabularx}
\end{table}

\begin{table}[htbp]
\caption{Results for $B_{s0}^{\ast\,0} \to f_0 X$ decays  
with $X=\pi^0,\;\eta^\prime,\;\eta,\;K^0,\;\omega,\;\rho^0$ .}
\label{tab:4}
\begin{tabularx}{\textwidth}{|l|Xr|Xr|Xr|}
\hline
&\multicolumn{2}{c|}{local limit}
&\multicolumn{2}{c}{$\Lambda_{B_{s0}^\ast}=2$ GeV, $\Lambda_{f_0}=1$ GeV}
&\multicolumn{2}{|c|}{$\Lambda_{B_{s0}^\ast}=1$ GeV, 
$\Lambda_{f_0}=1$ GeV}\\
\hline

Channel&$g_{B_{s0}^\ast f_0X}$&$\Gamma_{B_{s0}^\ast\to f_0 X}$ [GeV]
&$g_{B_{s0}^\ast f_0X}$&$\Gamma_{B_{s0}^\ast~\to~f_0 X}$ [GeV] 
&$g_{B_{s0}^\ast f_0X}$&$\Gamma_{B_{s0}^\ast~\to~f_0 X}$ [GeV]\\ 
\hline

$B_{s0}^{\ast\,0}\to f_0\pi^0$&$1.30\cdot10^{-8}$ GeV&$5.66\cdot10^{-19}$
&$5.43\cdot10^{-9}$ GeV&$9.93\cdot10^{-20}$&$3.86\cdot10^{-9}$ GeV
&$5.03\cdot10^{-20}$\\ 
\hline

$B_{s0}^{\ast\,0}\to f_0\eta^\prime$&$3.35\cdot10^{-8}$ GeV
&$3.67\cdot10^{-18}$&$1.43\cdot10^{-8}$ GeV&$6.69\cdot10^{-19}$
&$1.03\cdot10^{-8}$ GeV&$3.49\cdot10^{-19}$
\\
\hline

$B_{s0}^{\ast\,0}\to f_0\eta$&$<5.89\cdot10^{-9}$ GeV
&$<1.16\cdot10^{-19}$&$<2.48\cdot10^{-9}$ GeV&$<2.05\cdot10^{-20}$
&$<1.77\cdot10^{-9}$ GeV&$<1.05\cdot10^{-20}$
\\ 
\hline

$B_{s0}^{\ast\,0}\to f_0 K^0$&$4.19\cdot10^{-9}$ GeV&$5.88\cdot10^{-20}$
&$1.77\cdot10^{-9}$ GeV&$1.04\cdot10^{-20}$&$1.26\cdot10^{-9}$ GeV
&$5.32\cdot10^{-21}$\\ 
\hline

$B_{s0}^{\ast\,0}\to f_0\rho^0$&$5.89\cdot10^{-10}$&$4.64\cdot10^{-19}$
&$2.63\cdot10^{-10}$&$9.22\cdot10^{-20}$&$2.08\cdot10^{-10}$
&$5.75\cdot10^{-20}$\\ 
\hline

$B_{s0}^{\ast\,0}\to f_0\omega$&$6.69\cdot10^{-10}$&$5.86\cdot10^{-19}$
&$2.99\cdot10^{-10}$&$1.17\cdot10^{-19}$&$2.36\cdot10^{-10}$
&$7.31\cdot10^{-20}$\\ 
\hline
\end{tabularx}
\end{table}

\section{${\mathbf{D_s^+ \to f_0 \pi^+ decay}}$} \label{sec:3}

In this section we analyze the $D_s^+\to f_0\pi^+$ decay in order to 
derive a value for the $D^\ast K\pi$ coupling constant $g_\pi$. This 
coupling is needed for the calculation of the $D_{s1}\to f_0\pi$ decay 
width discussed in the next section. In this context we also obtain the 
decay width $\Gamma(D^\ast\to K\pi)$ as an additional result. 
\begin{figure}[htbp]
\includegraphics[trim= 0cm 16cm 0cm 7.0cm,clip,scale=0.6]{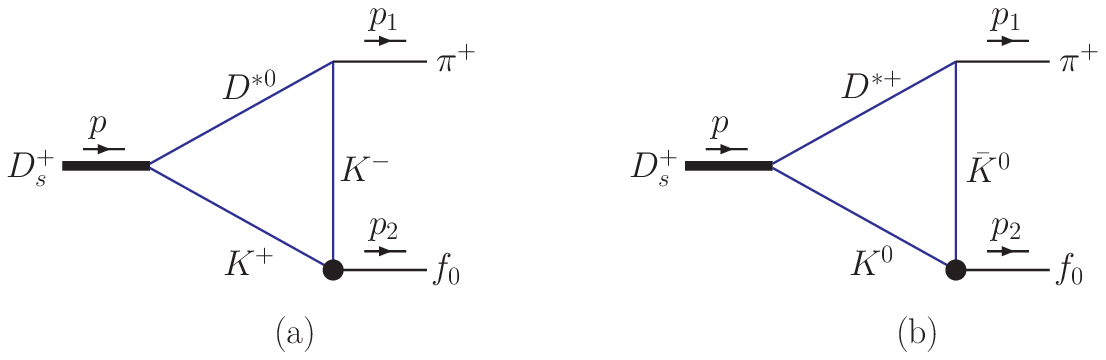}
\caption{$D_{s}$-decay.}
\label{fig4}
\end{figure}
The $D_s$-decay is illustrated by the Feynman diagrams of Fig.~\ref{fig4}, 
where the decay width is defined as 
\be
\Gamma(D_s^+\to f_0\pi^+) = 
\frac{g_{D_s f_0\pi}^2} 
{16\,\pi\,m_{D_s}^3} \, 
\lambda^\frac12(m_{D_s}^2,m_{f_0}^2,m_\pi^2) 
\label{eq:1}\,.
\en
The decay coupling
\bea
g_{D_s f_0\pi}&=&\frac{g_fg_{D_s}g_\pi}{(4\pi)^2}
\big[I(m_{D^{\ast\,+}}^2,m_{K^0}^2)+I(m_{D^{\ast\,0}}^2,m_{K^+}^2)\big]
\label{eq:2}
\ena
can be computed from the loop integral $I(m_{D^{\ast}}^2,m_{K}^2)$ 
given by
\be
I(m_{D^{\ast}}^2,m_{K}^2)&=&\int\frac{d^4k}{\pi^2i}
\,\widetilde \Phi_f(-k^2) \,
\big(p_{\pi}-k-\frac p2\big)_\mu \big(k-\frac p2-p_{D_s}\big)_\nu 
\,S_D^{\mu\nu}\big(k-\frac p2+p_{D_{s}}\big)S_K\big(k-\frac p2\big)
S_K\big(k+\frac p2\big)\,.
\en
The coupling constant $g_{D_s}$ of the $D_s D^\ast K$ interaction vertex 
has been estimated in two different QCD sum rule 
approaches~\cite{Wang:2006ida,Bracco:2006xf}, where both results do not 
vary significantly from each other. Here we use the result of the QCD sum 
rule approach in~\cite{Wang:2006ida} with $g_{D_s}=2.02$. 
By using the branching ratio 
$\text{Br}(D_s^+\to f_0\pi^+)=(6.0\pm2.4)\cdot10^{-3}$~\cite{Amsler:2008zz}, 
corresponding to $\Gamma(D_s^+\to f_0\pi^+)=7.9\cdot10^{-15}$~GeV, 
$g_\pi$ can be easily derived 
from (\ref{eq:1}) and (\ref{eq:2}): 
\eq 
g_\pi=6.41\cdot10^{-5}\,.\label{eq:3}
\en 
Now, the $D^\ast\to K\pi$ decay width is immediately given by 
\eq 
\Gamma(D^\ast\to K\pi)=
\frac{g_\pi^2}{48\pi m_{D^\ast}^5}
\, \lambda^\frac32(m_{D^\ast}^2,m_K^2.m_\pi^2) 
\en 
which leads to 
$\Gamma(D^{\ast} \to  {K}\pi) = 4.45\cdot10^{-11}\text{ GeV}\,.$ 

\section{${\mathbf{D_{s1}(2460)}}$ and 
${\mathbf{B_{s1}(5778)}}$ Decays}\label{sec:4}

\begin{figure}[htbp]
\includegraphics[trim= 0cm 16cm 0cm 7.0cm,clip,scale=0.6]{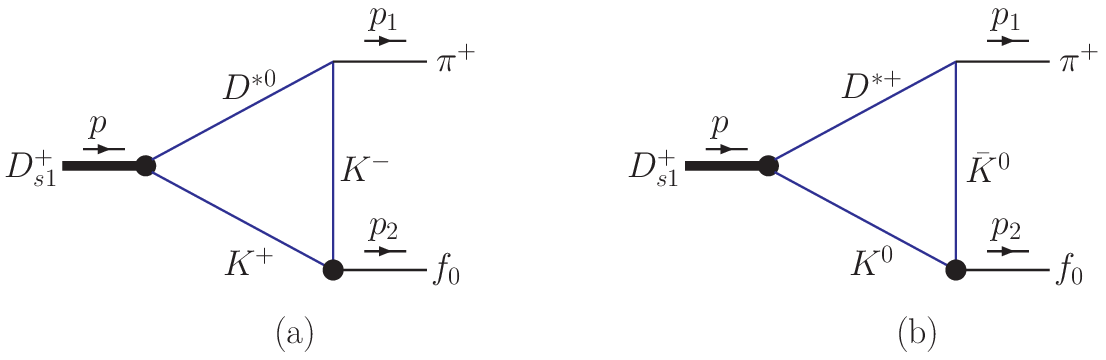}
\caption{$D_{s1}^\ast(2460)$ decay.}
\label{fig5}
\end{figure}

In this section we study the properties of the weak transitions between 
the axial vector hadronic molecules $D_{s1}(2460)$ and $B_{s1}(5778)$ and 
the scalar $f_0(980)$. The determination of $g_\pi$ in the last section 
enables us to compute the decay $D_{s1}^+(2460)\to f_0\pi^+$ within the 
$K$ $D^\ast$ bound state framework. The Feynman diagrams which contribute 
to this decay are illustrated in Fig.~\ref{fig5}.
In the first step we define the matrix element of the $D_{s1}^+\to 
f_0\pi^+$ transition in terms of the form factors $F_\pm$ and 
$p_\pm=p_f\pm p_\pi$
\eq 
{\cal M}^\mu&=&\frac{g_fg_{D_{s1}}g_\pi}{(4\pi)^2}
\big(I^\mu(m_{D^{\ast\,+}}^2,m_{K^0}^2)
+I^\mu(m_{D^{\ast\,0}}^2,m_{K^+}^2)\big)\nonumber\\
&=&F_+(m_{D_{s1}},m_{\pi},m_{f_0})p_+^\mu
+F_-(m_{D_{s1}},m_{\pi},m_{f_0})p_-^\mu\,,
\en 
where $p_f$ and $p_\pi$ are the $f_0$ and $\pi$ momenta, 
respectively. 

The loop integral involving the constituent kaons and $D^\ast$ meson 
is of the structure
\be
I^\mu(m_{D^{\ast}}^2,m_{K}^2)&=&\int\frac{d^4k}{\pi^2i}\,
\widetilde\Phi_{f_0}(-k^2)\widetilde\Phi_{D_s}
\big(-\big(k-\frac p2+\omega p_{D^\ast}\big)^2\big)
\,\big(p_\pi-k-\frac p2\big)_\nu \nonumber\\
&\times&S^{\mu\nu}_{D^\ast}\big(k-\frac p2+p_{D_{s1}}\big)
S_K\big(k-\frac p2\big)S_K\big(k+\frac p2\big)\,. 
\en 
The form factor $F_-$ defines the coupling 
$g_{D_{s1}f_0\pi} = F_-(m_{D_{s1}},m_{\pi},m_{f_0})$ 
which characterizes the decay width given by the expression
\bea
\Gamma(D_{s1}^+\to f_0\pi^+)&=&
\frac{g_{D_{s1}f_0\pi}^2}{48\pi\,m_{D_{s1}^+}^5} 
\, \lambda^\frac32(m_{D_{s1}^+}^2,m_{f_0}^2,m_{\pi^+}^2) 
\,. 
\ena

We compute the decay width for $D_{s1}^+\to f_0\pi^+$ for the $f_0$ 
size parameter $\Lambda_{f_0}$=1 GeV while $\Lambda_{D_{s1}}$ is 
varied between 1 GeV and 2 GeV. 

The results for the $D_{s1}\to f_0\pi$ decay width obtained within our 
hadronic molecule approach range from 
\eq 
\Gamma(D_{s1}\to f_0\pi)=2.85\cdot10^{-11}\text{ GeV},\quad 
\text{where} \quad g_{D_{s1}f_0\pi}=5.46\cdot 10^{-5}\quad\text{ at } 
\Lambda_{D_{s1}}=1 \text{ GeV}
\en 
to
\eq 
\Gamma(D_{s1}\to f_0\pi)=4.35\cdot 10^{-11}\text{ GeV},
\quad \text{where}\quad g_{D_{s1}f_0\pi}=6.74\cdot 10^{-5}
\quad\text{ at }\Lambda_{D_{s1}}=2 \text{ GeV.}
\en 

By analogy, we can also study the $B_{s1}\to f_0X$ decay, where $P$ 
represents a pseudoscalar final state. However, since no data are 
available to determine the $B^\ast f_0P$ coupling strength $g_{B^\ast}$, 
we quote the width and corresponding decay coupling in dependence on 
$g_{B^\ast}$. Varying $\Lambda_{B_{s1}}$ from 1.0 GeV to 2 GeV the 
width lies between
\eq 
\Gamma(B_{s1}\to f_0\pi)=8.82\cdot10^{-6}\, g_{B^\ast}^2
\text{ GeV,  where}\quad g_{B_{s1}f_0\pi}=0.016  \, g_{B^\ast} 
\quad \text{ at }\Lambda_{B_{s1}}=1 \text{ GeV}
\en 
and
\eq 
\Gamma(B_{s1}\to f_0\pi)=4.03\cdot10^{-5}\, g_{B^\ast}^2
\text{ GeV,  where}\quad g_{B_{s1}f_0\pi}=0.034 \, g_{B^\ast} 
\quad\text{ at }\Lambda_{B_{s1}}=2 \text{ GeV.}
\en 
  
\section{Summary}

In the present paper we focused on weak hadronic production processes 
of the scalar $f_0(980)$. For this purpose we studied the weak 
non-leptonic decays of the heavy mesons $D_{s0}^{\ast\,+}$, $D_{s1}^+$ 
as well as the $B_{s0}$ and $B_{s1}$ mesons assigned as the 
corresponding states in the bottom-strange sector. 

The formalism presented provides a clear and straightforward method to 
study the issue of hadronic molecules. Since all coupling constants are 
either fixed self-consistently by the compositeness condition or are 
deduced from experimental data, the only adaptive variables are the 
size parameters of the meson molecules which allow for their extended 
structure. Finite size effects are studied by varying the size 
parameters within a physically reasonable region between 1 and 2 GeV. 
Additionally we also compare the results with finite size effects to 
the local case related to point-like interactions. 

The molecular interpretation of both, the initial heavy mesons and the 
final decay product - the kaonic bound state $f_0$ - in the weak decays 
possibly offers a sensitive tool to study the structure issue. 
In particular for the $D_{s0}^\ast(2317)\to f_0X$ transitions we give 
clear predictions for the decay pattern arising in the hadronic molecule 
picture, both for $D_{s0}^\ast$ and $f_0$. Similarly, the result for the 
process $D_{s1}\to f_0 \pi$ is a straightforward consequence of the 
molecular interpretation. In addition the $D^\ast\to f_0\pi$ decay 
properties can also be used to get information on the $f_0$ substructure. 

Presently no comparative calculations, as for example in the full or 
partial quark-antiquark interpretation of the 
$D_{s0}^{\ast\, +},D_{s1}^+$ and $f_0$ mesons, exist. Hence, the real 
sensitivity of the results for the weak processes studied here on details 
of the meson structure remains to be seen. But judging from previous model 
calculations of for example the dominant observed decay modes of the 
$D^\ast_{s0}$ and $D_{s1}$ a strong dependence on the structure models 
can be expected. Therefore, upcoming experiments measuring the weak 
production processes involving the scalar meson $f_0(980)$ could lead to 
new insights into the meson spectrum and its structure issue. 

\begin{acknowledgments}

This work was supported by the DFG under Contracts No. FA67/31-1, 
No. FA67/31-2, and No. GRK683. This research is also part of the EU 
Integrated Infrastructure Initiative Hadronphysics project under 
Contract No. RII3-CT-2004-506078 and the President Grant of Russia
``Scientific Schools''  No. 817.2008.2. 

\end{acknowledgments}


\begin{thebibliography}{999}

\bibitem{Klempt:2007cp}
E.~Klempt and A.~Zaitsev,  
Phys.\ Rept.\  {\bf 454}, 1 (2007). 

\bibitem{Aubert:2003fg}
B.~Aubert {\em et~al.} ({\it BABAR} Collaboration),  
Phys. \ Rev. \ Lett. {\bf 90}, 242001 (2003). 

\bibitem{Besson:2003cp} 
D.~Besson {\em et~al.} (CLEO Collaboration),  
Phys. \ Rev. \ D {\bf 68}, 032002 (2003). 

\bibitem{Abe:2003jk} 
Y.~Mikami {\em et~al.} (Belle Collaboration), 
Phys. \ Rev. \ Lett. {\bf 92}, 012002 (2004). 

\bibitem{Barnes:2003dj}
  T.~Barnes, F.~E.~Close and H.~J.~Lipkin,
  Phys.\ Rev.\ D {\bf 68}, 054006 (2003). 

\bibitem{vanBeveren:2003kd}
  E.~van Beveren and G.~Rupp,
  Phys.\ Rev.\ Lett.\  {\bf 91}, 012003 (2003). 

\bibitem{Cheng:2003kg}
H.~Y.~Cheng and W.~S.~Hou, 
Phys. \ Lett. \ B {\bf 566}, 193 (2003). 

\bibitem{Godfrey:2003kg}
S.~Godfrey, 
Phys. \ Lett. \ B {\bf 568}, 254 (2003). 

\bibitem{Colangelo:2003vg}
P.~Colangelo and F.~De~Fazio,  
Phys. \ Lett. \ B {\bf 570}, 180 (2003). 

\bibitem{Bardeen:2003kt}
W.~A.~Bardeen, E.~J.~Eichten, and C.~T.~Hill, 
Phys. \ Rev. \ D {\bf 68}, 054024 (2003). 

\bibitem{Kolomeitsev:2003ac}
  E.~E.~Kolomeitsev and M.~F.~M.~Lutz,
  Phys.\ Lett.\  B {\bf 582}, 39 (2004)

\bibitem{Fayyazuddin:2003dp}
  Fayyazuddin and Riazuddin,
  Phys.\ Rev.\  D {\bf 69}, 114008 (2004). 

\bibitem{Ishida:2003gu}
  S.~Ishida, M.~Ishida, T.~Komada, T.~Maeda, M.~Oda,
  K.~Yamada and I.~Yamauchi, 
  AIP Conf.\ Proc.\  {\bf 717}, 716 (2004). 

\bibitem{Azimov:2004xk}
  Y.~I.~Azimov and K.~Goeke,
  Eur.\ Phys.\ J.\ A {\bf 21}, 501 (2004). 

\bibitem{Colangelo:2004vu}
P.~Colangelo, F.~De~Fazio, and R.~Ferrandes,  
Mod. \ Phys. \ Lett. \ A {\bf 19}, 2083 (2004). 

\bibitem{Mehen:2004uj}
  T.~Mehen and R.~P.~Springer,
  Phys.\ Rev.\  D {\bf 70}, 074014 (2004)

\bibitem{Hayashigaki:2004st}
  A.~Hayashigaki and K.~Terasaki,
  Prog.\ Theor.\ Phys.\  {\bf 114}, 1191 (2005). 

\bibitem{Colangelo:2005hv}
P.~Colangelo, F.~De~Fazio, and A.~Ozpineci, 
Phys. \ Rev. \ D {\bf 72}, 074004 (2005). 

\bibitem{Close:2005se}
F.~E.~Close and E.~S.~Swanson, 
Phys. \ Rev. \ D {\bf 72}, 094004 (2005). 

\bibitem{Wei:2005ag}
  W.~Wei, P.~Z.~Huang and S.~L.~Zhu,
  Phys.\ Rev.\ D {\bf 73}, 034004 (2006). 

\bibitem{Lu:2006ry}
  J.~Lu, X.~L.~Chen, W.~Z.~Deng and S.~L.~Zhu,
  Phys.\ Rev.\ D {\bf 73}, 054012 (2006). 

\bibitem{Rosner:2006vc}
  J.~L.~Rosner,
  Phys.\ Rev.\ D {\bf 74}, 076006 (2006). 

\bibitem{Swanson:2006st}
  E.~S.~Swanson,
  Phys.\ Rept.\  {\bf 429}, 243 (2006). 

\bibitem{Guo:2006fu} 
F.~K.~Guo, P.~N.~Shen, H.~C.~Chiang and R.~G.~Ping,
Phys.\ Lett.\  B {\bf 641}, 278 (2006); 
F.~K.~Guo, P.~N.~Shen, and H.~C.~Chiang,  
Phys. \ Lett. \ B {\bf 647}, 133 (2007). 

\bibitem{Liu:2006jx}
  X.~Liu, Y.~M.~Yu, S.~M.~Zhao and X.~Q.~Li,
  Eur.\ Phys.\ J.\  C {\bf 47}, 445 (2006). 

\bibitem{Wang:2006mf}
Z.~G.~Wang, 
Phys. \ Rev. \ D {\bf 75}, 034013 (2007). 

\bibitem{Gamermann:2007bm}
D.~Gamermann, L.~R. Dai, and E.~Oset, 
Phys. \ Rev. \ C {\bf 76}, 055205 (2007). 

\bibitem{Lutz:2007sk}
  M.~F.~M.~Lutz and M.~Soyeur,
  Nucl.\ Phys.\  A {\bf 813}, 14 (2008). 

\bibitem{Guo:2007up}
  F.~K.~Guo, S.~Krewald and U.~G.~Meissner,
  Phys.\ Lett.\  B {\bf 665}, 157 (2008). 

\bibitem{Guo:2008gp}
  F.~K.~Guo, C.~Hanhart, S.~Krewald and U.~G.~Meissner,
   Phys.\ Lett.\  B {\bf 666}, 251 (2008). 

\bibitem{Faessler:2007us}
A.~Faessler, T.~Gutsche, V.~E. Lyubovitskij, and Y.~L.~Ma,  
Phys. \ Rev. \ D {\bf 76}, 114008 (2007).  

\bibitem{Faessler:2007gv}
A.~Faessler, T.~Gutsche, V.~E. Lyubovitskij, and Y.~L.~Ma,  
Phys. \ Rev. \ D {\bf 76}, 014005 (2007). 

\bibitem{Faessler:2007hm}
A.~Faessler, T.~Gutsche, S.~Kovalenko and V.~E.~Lyubovitskij,
Phys.\ Rev.\ D {\bf 76}, 014003  (2007); 
A.~Faessler, T.~Gutsche and V.~E.~Lyubovitskij,
Prog.\ Part.\ Nucl.\ Phys.\  {\bf 61}, 127 (2008). 

\bibitem{Faessler:2008vc}
A.~Faessler, T.~Gutsche, V.~E. Lyubovitskij, and Y.~L.~Ma,  
Phys. \ Rev. \ D {\bf 77}, 114013 (2008). 

\bibitem{Dong:2008gb}
Y.~B.~Dong, A.~Faessler, T.~Gutsche and V.~E.~Lyubovitskij,
Phys.\ Rev.\  D {\bf 77}, 094013 (2008). 

\bibitem{Branz:2007xp}
T.~Branz, T.~Gutsche, and V.~E.~Lyubovitskij, 
Eur. \ Phys. \ J. \ A {\bf 37}, 303 (2008). 

\bibitem{Branz:2008ha}
T.~Branz, T.~Gutsche, and V.~E. Lyubovitskij,
Phys. \ Rev. \ D {\bf 78}, 114004 (2008).  

\bibitem{Weinberg:1962hj}
S.~Weinberg, Phys. \ Rev. {\bf 130}, 776 (1963). 

\bibitem{Salam:1962ap}
A.~Salam, Nuovo \ Cim. {\bf 25}, 224 (1962).

\bibitem{Efimov:1993zg}
G.~V.~Efimov and M.~A.~Ivanov, 
{\it The Quark Confinement Model of Hadrons} 
(IOP Publishing, Bristol $\&$ Philadelphia, 1993).

\bibitem{Ivanov:1996pz} 
M.~A.~Ivanov, M.~P.~Locher and V.~E.~Lyubovitskij, 
Few Body Syst.\  {\bf 21}, 131 (1996); 
M.~A.~Ivanov, V.~E.~Lyubovitskij, J.~G.~K\"orner and P.~Kroll,
Phys.\ Rev.\ D {\bf 56}, 348 (1997); 
M.~A.~Ivanov, J.~G.~K\"orner, V.~E.~Lyubovitskij and A.~G.~Rusetsky, 
Phys.\ Rev.\ D {\bf 60}, 094002 (1999); 
A.~Faessler, T.~Gutsche, M.~A.~Ivanov, J.~G.~K\"orner  
and V.~E.~Lyubovitskij, 
Phys.\ Lett.\ B {\bf 518}, 55 (2001); 
A.~Faessler, T.~Gutsche, M.~A.~Ivanov, J.~G.~K\"orner,
V.~E.~Lyubovitskij, D.~Nicmorus and K.~Pumsa-ard,
Phys.\ Rev.\ D {\bf 73}, 094013 (2006);  
A.~Faessler, T.~Gutsche, B.~R.~Holstein, V.~E.~Lyubovitskij, 
D.~Nicmorus and K.~Pumsa-ard, 
Phys.\ Rev.\  D {\bf 74}, 074010 (2006); 
A.~Faessler, T.~Gutsche, B.~R.~Holstein, M.~A.~Ivanov, 
J.~G.~Korner and V.~E.~Lyubovitskij, 
Phys.\ Rev.\  D {\bf 78}, 094005 (2008). 

\bibitem{Baru:2003qq}
V.~Baru, J.~Haidenbauer, C.~Hanhart, Yu.~Kalashnikova, and A.~E.~Kudryavtsev,
  Phys.\ Lett.\  B {\bf 586} (2004) 53. 

\bibitem{Hanhart:2007wa}
C.~Hanhart, Y.~S.~Kalashnikova, A.~E.~Kudryavtsev, and A.~V.~Nefediev, 
Phys. \ Rev. \ D {\bf 75}, 074015 (2007). 

\bibitem{Amsler:2008zz}
C.~Amsler {\em et~al.} (Particle Data Group), 
Phys. \ Lett. \ B {\bf 667}, 1 (2008).

\bibitem{Guo:2008zg}
  F.~K.~Guo, C.~Hanhart and U.~G.~Meissner,
  Phys.\ Lett.\  B {\bf 665}, 26 (2008). 

\bibitem{Eidelman:2004wy}
S.~Eidelman {\em et~al.} (Particle Data Group), 
Phys. \ Lett. \ B {\bf 592}, 1 (2004).

\bibitem{Wang:2006ida}
Z.~G.~Wang and S.~L.~Wan,
Phys. \ Rev. \ D {\bf 74}, 014017 (2006). 

\bibitem{Bracco:2006xf}
M.~E. Bracco, A.~Cerqueira Jr., M.~Chiapparini, A.~Lozea, and M.~Nielsen,
Phys. \ Lett. \ B {\bf 641}, 286 (2006). 

\end{thebibliography}
\end{document}